# Local Filtering Fundamentally Against Wide Spectrum

Arising from C. C. Chen *et al*. Nature 496, 74-79 (4 April 2013)

**Chen *et al*. [1]** *applied "three-dimensional (3D) Fourier filtering together with equal-slope tomographic reconstruction" for an observation of "nearly all the atoms in a multiply twinned platinum nanoparticle"*. However, their methodology suffers from fundamental methodological flaws, as initially brought up by a recent Communications Arising [2] and now analyzed in-depth as follows.

Our analysis is significantly different from that in [2], and builds a direct and definite case against the work presented in [1]. It was correctly reasoned in [2] that *"It could be that essential diffuse scattering is lost in the noise… It does not necessarily put atoms in the right places because essential Fourier components are missing.... Even when the noise threshold is set to 10%, there are still considerable displacements."* Namely, information fidelity is compromised due to "*suppressing essential diffuse contributions*" [2]. On the other hand, here we rely on the fact that both dislocation phenomena and noise background have broad Fourier spectra, and the proposed Fourier filtering method around the Bragg spots [1] is fundamentally restricted from capturing the wide spreadability of both dislocation and noise in the Fourier space.

While in [2] the authors pointed out substantial displacements of atoms, here in Figure 1 we show that the local Fourier filtering method used in [1] can completely miss atomic disarrangement and/or produce ghost features that resemble atomic positions. It is underlined that we have made our case using the same methodology, data and parameters as those reported in [1] and very recently shared in [3]. The two counter-examples in Figure 1 are equivalent to the commonly known Type I and Type II errors: (1) false alarm ("*False Atom-like Structures*") and (2) missing target ("*Missing Disarrangement*") respectively. The underlying common cause of both the problems is the major mismatch between a limited mask size required by the local Fourier filtering method and a much larger spectral support of either a sharp abnormality or a noisy background.

In response to [2], Dr Miao's group argued in [4] that *"...if all we did was simple Fourier filtering with small apertures around the Bragg spots, then this would indeed lead to artefacts; we avoided this by verifying results against unbiased Wiener filters as well as by using relatively large apertures which were adjusted to minimize signal loss".* However, their argument is invalid. If the traditional Wiener filtering method can validate the proposed filtering method, why do we need the latter? Clearly, the new method was supposed to be of sufficiently higher quality than the Wiener filtering method. Clearly, the neighborhood of the Bragg spots is intrinsically localized, and the proposed filtering scheme is intended to match the locality. When simple Fourier filtering is applied with small or even relatively large apertures around the Bragg spots, what is excluded is not only high-frequency noise but also potentially a significant portion of high-frequency components needed for quantification of dislocations. On the other hand, noise contained in an array of filtering apertures could generate significant artefacts giving an illusion of arranged atoms as mentioned in [2]. While an optimal aperture size remains either unknown (in this regard, no specific guideline was given in [1], and all possibilities are open in challenging cases) or non-existent at all, the risk of introducing Type I and II errors is realistic and quite likely, as shown in Figure 1.

On 27 May 2014, we received Dr. Miao's rebuttal in which he criticized our 2D example presented in an initial communication. He wrote that "*our Fourier filter is three-dimensional (3D), but theirs is two-dimensional (2D).*" "*3D Fourier filtering is more accurate than the 2D case. Furthermore, we optimized our 3D Fourier filter using multislice simulation data*". As a matter of fact, in the Fourier reconstruction context there is no essential difference between 2D and 3D cases, as evidenced by Figure 1 which was produced after carefully following Dr. Miao's MatLab instructions shared in [3] including the aforementioned multislice simulation and other details. Indeed, we have reproduced Dr. Miao's key images in [1] but we have found more problems with their program. In their Wiener filtering process, the noise power spectrum was estimated from the original image by averaging its Fourier spectrum over spherical shells respectively. Such averages assumed *no structure* in the object and spherically symmetric noise characteristics, which is generally invalid.

In summary, our primary concern is not only general but also specific that the particular results reported in [1] are subject to Type I and II errors. To address this local filtering issue, we suggest to use an advanced iterative algorithm such as the dictionary learning method described in [5], which avoids the fundamental problems associated with the local filtering method in [1] whereas the so-called equal-slope reconstruction has little advantage in this scenario [5].

Ge Wang[1], Hengyong Yu[2], Scott S. Verbridge[3], Lizhi Sun[4]

1. Biomedical Imaging Center, Department of Biomedical Engineering, Rensselaer Polytechnic Institute, Troy, NY 12180, USA
2. Department of Biomedical Engineering, Wake Forest University, Winston-Salem, NC 27157, USA
3. School of Biomedical Engineering & Sciences, Virginia Tech, Blacksburg, VA 24061, USA
4. Departments of Civil & Environmental Engineering and Chemical Engineering & Materials Science, University of California, Irvine, CA 92697, USA

ge-wang@ieee.org, hyu@wakehealth.edu, sverb@vt.edu, lsun@uci.edu

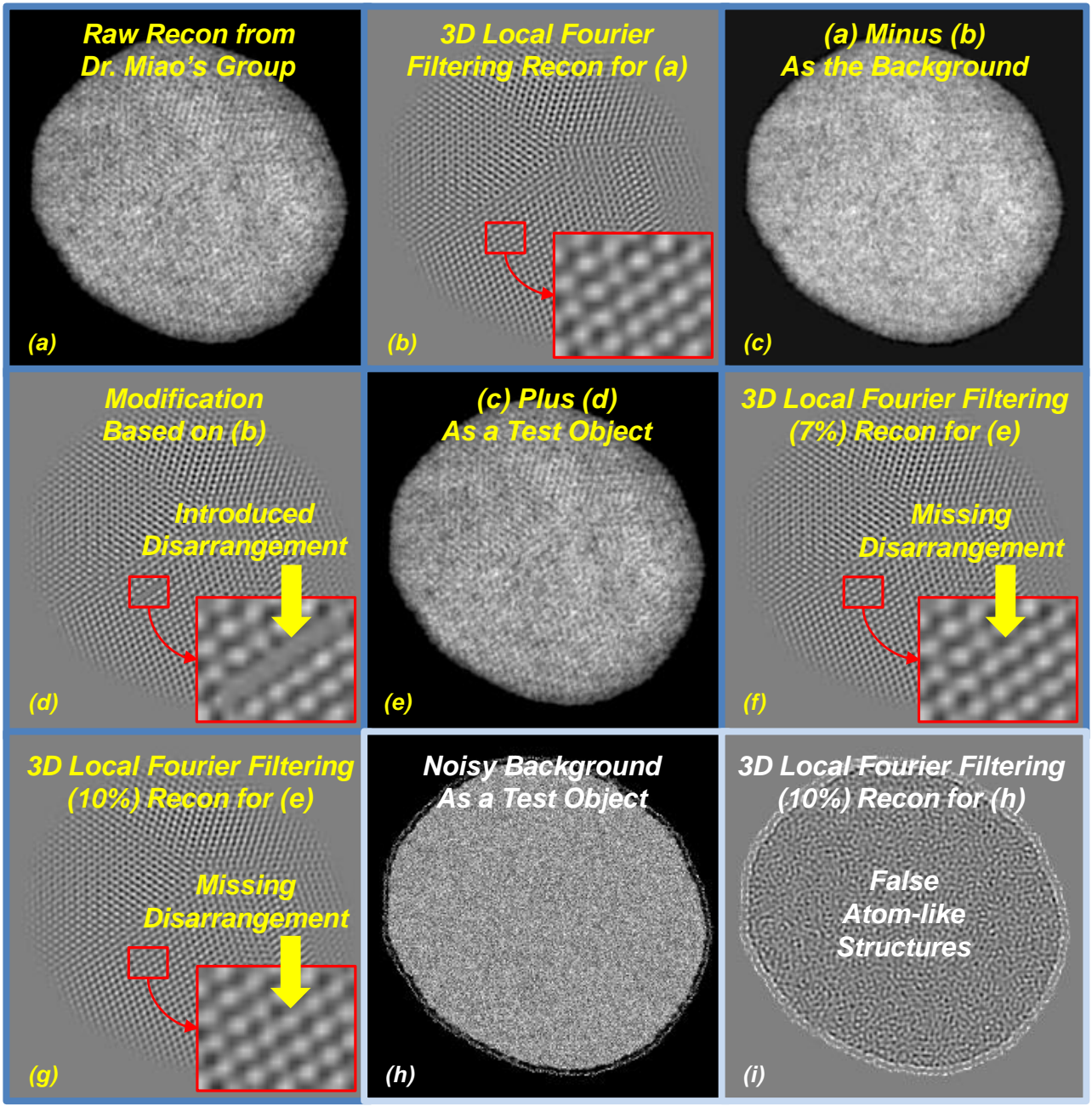

**Figure 1. Fundamental flaw of the local Fourier filtering method described in [1]. (a)** A 2.6-angstrom-thick slice through the x-y plane of a raw 3D reconstruction with the z-axis along the beam direction (from Dr. Miao's website www.physics.ucla.edu/research/imaging/dislocations); **(b)** the same slice of the 3D structure after a 3D local Fourier filtration (with 7% cutoff); **(c)** the difference between (a) and (b) (which is the background discarded by the local filtering method); **(d)** a modified version of (b) after removal of several atoms in the 3D structure to form a linear disarrangement; **(e)** the combination of (c) and (d) into a new 3D test object which is the same as (a) except for the inclusion of the sharp dislocation in (d); **(f)** the image reconstructed for the 3D test object shown in (e) using the local filtering method (7% cutoff), which contains no abnormality when it actually exists; **(g)** a variant of (f) (10% cutoff), which shows the same problem; **(h)** a 3D noisy background within the object support through the x-y plane (positive elements set to 1, the rest to 0, and a Gaussian noise distribution multiplied by the binary mask); and **(i)** the image reconstructed for the noisy background shown in (h) using the local filtering method (10% cutoff), generating many atom-like structures that make no sense.